\documentclass[prd,nofootinbib,showpacs]{revtex4}
\usepackage{natbib}
\usepackage{amssymb,amsbsy,amsmath,amsfonts}
\usepackage{graphicx}
\usepackage{float}
\usepackage{rotating}

\def\slashp{p \!\!\! \slash}

\def\slashD{D \!\!\! \slash}

\begin{document}
\title {SU(3)-breaking corrections to the hyperon vector coupling $f_1(0)$ in covariant baryon 
chiral perturbation theory}
 \author{L.S. Geng}
\author{J. Martin Camalich}
 \author{M.J. Vicente Vacas}
 \affiliation{
Departamento de F\'{\i}sica Te\'orica and IFIC, Universidad de
Valencia-CSIC, E-46071 Valencia, Spain}

\begin{abstract}
We calculate the SU(3)-breaking corrections to the hyperon vector coupling $f_1(0)$ up to $\mathcal{O}(p^4)$ in covariant baryon chiral perturbation theory with dynamical octet and decuplet contributions. We find that the decuplet contributions are
of similar or even larger size than the octet ones. Combining both,
we predict positive SU(3)-breaking corrections to all the four independent $f_1(0)$'s (assuming isospin symmetry), which are
consistent, within uncertainties, with the latest results form large $N_c$ fits, chiral quark models, and quenched lattice QCD calculations.
\end{abstract}
\date{\today}
\maketitle
\section{Introduction}
Hyperon semileptonic decays, parameterized by three vector transition form factors ($f_1$, $f_2$, and $f_3$) and three axial form factors ($g_1$, $g_2$, and $g_3$), have received renewed interest in
recent years due to various reasons. In particular,  
they provide an alternative source~\cite{Cabibbo:2003cu,Cabibbo:2003ea,FloresMendieta:2004sk,Mateu:2005wi} 
to allow one to extract the Cabibbo-Kobayashi-Maskawa (CKM) 
matrix element $V_{us}$ ~\cite{Cabibbo:1963yz,Kobayashi:1973fv}, 
in addition to kaon semileptonic decays (see, e.g., Ref.~\cite{Blucher:2005dc} for a recent review),
hadronic decays of the $\tau$ lepton~\cite{Gamiz:2007qs} and the ratio $\Gamma(K^+\rightarrow \mu^+\nu_\mu)/\Gamma(\pi^+\rightarrow\mu^+\nu_\mu)$~\cite{Marciano:2004uf}. 
The hyperon vector coupling $f_1(0)$ plays an essential role in order to extract $V_{us}$ accurately.

Due to the Conservation of Vector Current (CVC) $f_1(0)$ is
known up to SU(3)-breaking effects, which are of subleading-order according to the Ademollo-Gatto theorem~\cite{Ademollo:1964sr}. Theoretical estimates of SU(3)-breaking corrections to $f_1(0)$ have
been performed in various frameworks, including quark models~\cite{Donoghue:1986th,Schlumpf:1994fb,Faessler:2008ix},
 large-$N_c$ fits~\cite{FloresMendieta:2004sk}, and
chiral perturbation theory (ChPT)~\cite{Krause:1990xc,Anderson:1993as,Kaiser:2001yc,Villadoro:2006nj,Lacour:2007wm}.
These SU(3)-breaking corrections have also been studied recently in quenched lattice QCD (LQCD) calculations  for
two of the four independent channels (assuming isospin symmetry): $\Sigma^-\rightarrow n$~\cite{Guadagnoli:2006gj} and $\Xi^0\rightarrow\Sigma^+$~\cite{Sasaki:2008ha}.

In principle, ChPT provides a model independent way to
estimate the SU(3)-breaking corrections to $f_1(0)$.  However, it is known that ChPT calculations 
converge slowly in SU(3) flavor space. This problem becomes even more pronounced
in the one-baryon sector, where the physics at a certain order can be blurred
by the power-counting restoration procedures, as can be clearly seen in
the case of the baryon octet magnetic moments~\cite{Geng:2008mf}. 
Fortunately, in the case of  $f_1(0)$,
the Ademollo-Gatto theorem dictates that up to $\mathcal{O}(p^4)$ no unknown LEC's contribute and, therefore, no
power-counting-breaking terms appear. Consequently, up to this order there is no need to apply any
power-counting restoration procedures and
a ChPT calculation is fully predictive.

In a recent $\mathcal{O}(p^4)$ calculation performed in Heavy Baryon (HB) ChPT, it was shown that the chiral series with only the octet contributions converge slowly  while the convergence is
completely spoiled by the inclusion of the decuplet ones~\cite{Villadoro:2006nj}.
In a later work~\cite{Lacour:2007wm}, the infrared version of baryon chiral perturbation theory (IRChPT)~\cite{Becher:1999he} was employed and
calculations were performed up to $\mathcal{O}(p^4)$ with only
the octet contributions. The slow convergence of the chiral series was confirmed but the importance of relativistic corrections was stressed.

In the present work, we perform the first covariant baryon ChPT calculation of  $f_1(0)$ 
up to $\mathcal{O}(p^4)$, including both octet and decuplet contributions. 
This article is organized as follows. In Sec. 2, we fix our notation and
write down the relevant chiral Lagrangians up to $\mathcal{O}(p^4)$. To study the contributions of the decuplet baryons, we adopt the ``consistent'' coupling scheme for the
Rarita-Schwinger description of the spin-3/2 decuplet fields~\cite{Pascalutsa:2000kd}. 
In Sec. 3,
we present our numerical results order by order, contrast them with the corresponding HBChPT and IRChPT results,
and study the convergence of the chiral series. We also compare our full results with those obtained in
other approaches, including large $N_c$ fits, quark models, and lattice QCD calculations. Finally, we use our results of $f_1(0)$ to extract $V_{us}$ from the experimental values of the decay rates and $g_1(0)/f_1(0)$. 
Summary and conclusions follow in Sec. 4.

\section{Formalism}

The baryon vector form factors as probed by the charged $\Delta$S=1 weak current
$V^\mu=V_{us}\bar{u}\gamma^\mu s$
are defined by
\begin{equation}
\langle B'\vert V^\mu\vert B\rangle =V_{us}\bar{u}(p')\left[\gamma^\mu f_1(q^2)+\frac{2i \sigma^{\mu\nu}q_\nu}{M_{B'}+M_B}f_2(q^2)+\frac{2 q^\mu }{M_{B'}+M_B}f_3(q^2)\right]u(p), 
\end{equation}
where $q=p'-p$. In the SU(3)-symmetric limit, $f_1(0)$ is fixed by the conservation of the SU(3)$_V$-charge $g_V$. Furthermore, the Ademollo-Gatto theorem states that  SU(3)-breaking corrections start at second order in the expansion parameter $m_s-m$
\begin{equation}\label{eq:ag}
f_1(0)=g_V+\mathcal{O}((m_s-m)^2),
\end{equation}
where $m_s$ is the strange quark mass and $m$ is the mass of the light quarks.
The values of $g_V$ are $-\sqrt{\frac{3}{2}}$, $-\frac{1}{\sqrt{2}}$, $-1$,  $\sqrt{\frac{3}{2}}$, $\frac{1}{\sqrt{2}}$,  $1$ for  $\Lambda\rightarrow p$, $\Sigma^0\rightarrow p$, 
 $\Sigma^-\rightarrow n$,  $\Xi^-\rightarrow \Lambda$, $\Xi^-\rightarrow \Sigma^0$, and 
$\Xi^0\rightarrow\Sigma^+$, respectively.
In the isospin-symmetric limit only four of these channels, which we take as $\Lambda\rightarrow N$, $\Sigma\rightarrow N$, $\Xi\rightarrow\Lambda$, and $\Xi\rightarrow\Sigma$, provide independent information  We will parameterize the SU(3)-breaking corrections order-by-order in the relativistic chiral expansion as follows: 
\begin{equation}
f_1(0)=g_V\left( 1+\delta^{(2)}+\delta^{(3)}+\cdots\right) \label{eq:Adem-Gatt},
\end{equation}
where $\delta^{(2)}$ and $\delta^{(3)}$ are the leading and next-to-leading order SU(3)-breaking corrections
induced by loops, corresponding to $\mathcal{O}(p^3)$ and $\mathcal{O}(p^4)$ chiral calculations.
\subsection{Chiral Lagrangians involving only octet baryons and pseudoscalars}
The lowest-order SU(3) chiral Lagrangian describing the pseudo-Goldstone bosons in the presence of an external vector current is:
\begin{equation}
\mathcal{L}^{(2)}_\phi=\frac{F_0^2}{4}\langle\nabla_\mu U(\nabla^\mu U)^\dagger+U\chi^\dagger+\chi U^\dagger\rangle, \label{eq:MesonLag}
\end{equation}
where the parameter $F_0$ is the chiral-limit  decay constant, $U$ is the SU(3) representation of the meson fields and $\nabla_\mu$ is its covariant derivative: $\nabla_\mu U=\partial_\mu -i [v_\mu,U]$,  with $v_\mu$ being the vector source. The explicit breaking of chiral symmetry comes from $\chi=2B_0\mathcal{M}$ where $B_0$ measures the strength of the  breaking and $\mathcal{M}={\rm diag}(m,m,m_s)$ is the quark mass matrix in the isospin symmetric limit~\cite{Scherer:2002tk}. In the above and forthcoming Lagrangians, the symbol $\langle ...\rangle$ denotes the trace in SU(3)
flavor space.

The lowest-order chiral Lagrangian describing octet baryons interacting with
pseudoscalars and an external vector source reads:
\begin{equation}
\mathcal{L}^{(1)}_{\phi \mathrm{B}}=\langle\bar{B}\left(i\slashD-M_0\right)B\rangle+\frac{D/F}{2}\langle\bar{B}\gamma^\mu\gamma_5[ u_\mu,B]_\pm\rangle, \label{eq:BaryonLag1}
\end{equation}
where $B$ denotes the traceless flavor matrix accounting for the octet-baryon fields, $M_0$ is the chiral-limit octet-baryon mass, $D$ and $F$ are the axial and vector meson-baryon couplings  and $D_\mu B=\partial_\mu B+[\Gamma_\mu,B]$ is the covariant derivative.  Furthermore, with $u^2\equiv U$, $u_\mu$ and $\Gamma_\mu$ are the so-called {\it vielbein} and {\it connection}:
\begin{eqnarray}\label{eq:VA}
u_\mu&=&i(u^\dagger\partial_\mu u-u\partial_\mu u^\dagger)+(u^\dagger v_\mu u-u v_\mu u^\dagger), \nonumber\\
\Gamma_\mu&=&\frac{1}{2}(u^\dagger\partial_\mu u+u\partial_\mu u^\dagger)-\frac{i}{2}(u^\dagger v_\mu u+u v_\mu u^\dagger).\label{eq:vielbeinConnection}
\end{eqnarray}

The only higher-order chiral Lagrangian that also contributes is through
the SU(3)-breaking of the masses of octet baryons\footnote{We have omitted a singlet term indistinguishable from $M_0$ for our purposes.}
\begin{equation}
\mathcal{L}^{(2)}_{\phi B}=b_{D/F}\langle\bar{B}[\chi_+,B]_\pm\rangle\nonumber
\end{equation}
with 
\begin{equation}
\chi_+ = 2\chi=4B_0\mathcal{M}=2{\rm diag}(m_\pi^2,m_\pi^2,2m^2_K-m^2_\pi),
\end{equation}
which leads to the following octet baryon masses up to this order:
\begin{eqnarray}\label{eq:mso}
  M_N =M_0+4 m_K^2 b_D-4(m_K^2-m_\pi^2)b_F, &\;\;&  M_\Sigma = M_0+4 m_\pi^2 b_D, \nonumber \\
  M_\Xi = M_0+4 m_K^2 b_D+4(m_K^2-m_\pi^2)b_F,  &\;\;& M_\Lambda =M_0+\frac{4}{3}  (4m^2_K-m^2_\pi)b_D .
\end{eqnarray}
Fitting the above masses to their corresponding physical values, with $m_\pi=0.138$ GeV, $m_K=0.496$ GeV,
and $m_\eta=0.548$ GeV, yields $M_0=1.197$ GeV, $b_D=-0.0661$ GeV$^{-1}$, and $b_F=0.2087$ GeV$^{-1}$,
which correspond to $M_N=0.942$(0.939) GeV, $M_\Sigma=1.192$(1.193) GeV, $M_\Xi=1.321$(1.318) GeV, and
$M_\Lambda=1.112$(1.116) GeV, with the physical values given in parentheses. It is clear that
the differences between the second order fits and the physical values are quite small. Using either of them will be numerically equivalent. In the $\mathcal{O}(p^4)$ calculation, we will use
the second order fits, Eq.~(\ref{eq:mso}), to keep track of the SU(3)-breaking pattern. While at
$\mathcal{O}(p^3)$, we use the average mass of the octet baryons, $M_B=1.151$ MeV, without introducing mass splittings.

\subsection{Chiral Lagrangians involving decuplet baryons}
In this work, we adopt the so-called ``consistent'' couplings~\cite{Pascalutsa:2000kd}  to describe
the interactions between
the decuplet and the octet baryons. Compared to
conventional couplings (see, e.g., Refs.~\cite{Pascalutsa:2000kd,Hacker:2005fh}), the consistent couplings
are more stringent due to the requirement that  all interactions have the same type of gauge
invariance as the kinetic term of the spin-3/2 fields~\cite{Pascalutsa:2000kd}. 
To calculate
$f_1(0)$ up to $\mathcal{O}(p^3)$, one only needs the following lowest-order chiral Lagrangians~\cite{{Geng:2009hh}}:
\begin{equation}
 \mathcal{L}^{(1)}_\mathrm{DD}=\bar{T}_\mu(\gamma^{\mu\nu\alpha}iD_\alpha-M_{D0}\gamma^{\mu\nu}) T_\nu,
\end{equation}
\begin{equation}\label{eq:consistent}
\mathcal{L}^{(1)}_\mathrm{DB}=\frac{i\mathcal{C}}{m_D}(D^\dagger_\mu \bar{T}_\nu\gamma^{\mu\nu\lambda} u_\lambda B+
\bar{B}u_\lambda\gamma^{\mu\nu\lambda}D_\mu T_\nu),
\end{equation}
where $M_{D0}$ is the chiral-limit decuplet-baryon mass, $D^\alpha T^\nu_{abc}=\partial^\alpha T^\nu_{abc}+(\Gamma^\alpha)^d_a T^\nu_{dbc}
+(\Gamma^\alpha)^d_b T^\nu_{adc}+(\Gamma^\alpha)^d_c T^\nu_{abd}$,
$T^\nu=T_{ade}\psi^\nu$, $\bar{T}^\mu=\bar{T}^{ade}\bar{\psi}^\mu$ with
the following associations:
$T_{111}=\Delta^{++}$, $T_{112}=\Delta^+/\sqrt{3}$,
$T_{122}=\Delta^0/\sqrt{3}$, $T_{222}=\Delta^-$, $T_{113}=\Sigma^{*+}/\sqrt{3}$,
$T_{123}=\Sigma^{*0}/\sqrt{6}$, $T_{223}=\Sigma^{*-}/\sqrt{3}$, 
$T_{133}=\Xi^{*0}/\sqrt{3}$, $T_{233}=\Xi^{*-}/\sqrt{3}$, and $T_{333}=\Omega^-$.
The value of the pseudoscalar-baryon-decuplet coupling $\mathcal{C}$ is determined
to be $\mathcal{C}\approx1.0$ from the $\Delta\rightarrow\pi N$ decay width.\footnote{Note the definition of $u_\mu$ in Eq.~(\ref{eq:VA}) is a factor of 2
different from that of HBChPT in Refs.~\cite{Villadoro:2006nj,Butler:1992pn}.} 
In SU(3) flavor space, the value of $\mathcal{C}$ can be different for different channels. In the present work,
as in Ref.~\cite{Villadoro:2006nj}, we use the same $\mathcal{C}$ for all the channels, assuming that SU(3)-breaking corrections to $f_1(0)$ induced by using channel-specific $\mathcal{C}$'s are of higher order.
The spin-3/2 propagator in $d$ dimensions is  
\begin{equation}
S^{\mu\nu}(p)=-\frac{\slashp+M_D}{p^2-M_D^2+i\epsilon}\left[
g^{\mu\nu}-\frac{1}{d-1}\gamma^\mu\gamma^\nu-\frac{1}{(d-1)M_D}(\gamma^\mu p^\nu-\gamma^\nu p^\mu)
-\frac{d-2}{(d-1)M_D^2} p^\mu p^\nu \right]
\end{equation}
with $M_D$ the decuplet baryon mass.

To calculate $f_1(0)$ at $\mathcal{O}(p^4)$, the following second-order chiral
Lagrangian is needed to break the mass degeneracy of the decuplet baryons\footnote{As
in the octet case, we have omitted a singlet term indistinguishable from $M_{D0}$ for our purposes.}
\begin{equation}
 \mathcal{L}^{(2)}_\mathrm{DD}=\frac{\gamma_M}{2} \bar{T}^\mu \chi_+ T_\mu,
\end{equation}
which leads to
\begin{eqnarray}
  M_\Delta= M_{D0} + 3 m_\pi^2 \gamma_M, &\;\;&
  M_{\Sigma^*} = M_{D0} + (2 m_K^2+m_\pi^2) \gamma_M\nonumber\\
  M_{\Xi^*} = M_{D0} + (4 m_K^2-m_\pi^2)\gamma_M, &\;\;&
  M_{\Omega}= M_{D0}+ 3 (2 m_K^2-m_\pi^2) \gamma_M.
\end{eqnarray}
A fit to the decuplet baryon masses, with the meson masses given above, yields $\gamma_M=0.3236$ GeV$^{-1}$ and
$m_{D0}=
1.216$ GeV, which correspond to $M_\Delta=1.235$(1.232) GeV, $M_{\Sigma^*}=1.382$(1.384) GeV, $M_{\Xi^*}=1.529$(1.533) GeV, and $M_\Omega=1.676$(1.672) GeV. As in the
octet case,  we use the second order fits in our calculation of the $\mathcal{O}(p^4)$ results,
while in the $\mathcal{O}(p^3)$ calculation, we use the average of the
decuplet baryon masses, $M_D=1.382$ GeV, for all the decuplet baryons.

\section{Results and discussions}
\subsection{SU(3)-breaking corrections to $f_1(0)$ due to octet contributions up to $\mathcal{O}(p^4)$}

\begin{figure}[t]
\includegraphics[scale=0.8,angle=270]{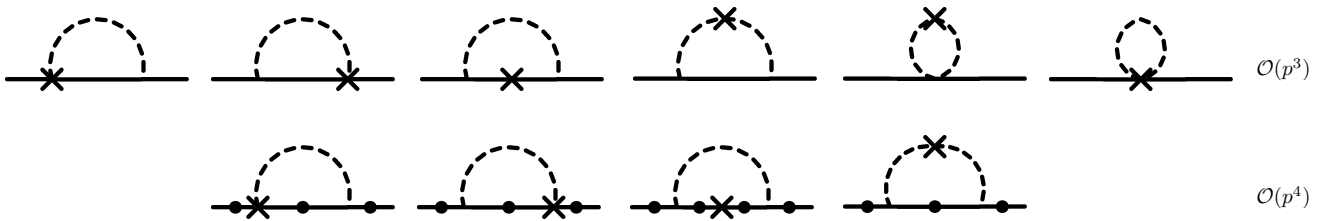}
\caption{Feynman diagrams contributing to the SU(3)-breaking corrections to 
the hyperon vector coupling $f_1(0)$
up to $\mathcal{O}(p^4)$. The solid lines correspond to baryons and dashed lines to mesons; crosses indicate the coupling of the external current; black dots denote mass splitting 
insertions. We have not shown explicitly those diagrams corresponding to wave function renormalization, which have been taken into account in the calculation.
\label{fig:doctet}}
\end{figure}
All the diagrams contributing to $f_1(0)$ up to $\mathcal{O}(p^4)$ are shown in
Fig.~\ref{fig:doctet}, where the leading and next-to-leading order SU(3)-breaking corrections
are given by the diagrams in the first and second row, respectively.

The $\mathcal{O}(p^3)$ results are quite compact and have the following
structure for the transition $i\rightarrow j$:
\begin{eqnarray}\label{eq:sumo3}
 \delta_B^{(2)}(i\rightarrow j)&=&\sum_{M=\pi,\eta,K}\beta^\mathrm{BP}_M H_\mathrm{BP}(m_M)+\sum_{M=\pi,\eta} \beta^\mathrm{MP}_{M} H_\mathrm{MP}(m_M,m_K)+\sum_{M=\pi,\eta,K} \beta^\mathrm{KR}_M H_\mathrm{KR}(m_M)\nonumber\\
&&-\frac{3}{8}\sum_{M=\pi,\eta}H_\mathrm{TD1}(m_M,m_K)+\frac{3}{8}\sum_{M=\pi,\eta}H_\mathrm{TD2}(m_M)
+\frac{3}{4}H_\mathrm{TD2}(m_K)\nonumber\\
&&+\frac{1}{2}\sum_{M=\pi,\eta,K}(\beta^\mathrm{WF}_M(i)+\beta^\mathrm{WF}_M(j))H_\mathrm{WF}(m_M),
\end{eqnarray}
where $\beta^\mathrm{BP}$, $\beta^\mathrm{MP}$, $\beta^\mathrm{KR}$, and $\beta^\mathrm{WF}$ are given in Tables \ref{table:o3bp}, \ref{table:o3mp}, \ref{table:o3kr}, and \ref{table:o3wf} in the Appendix, and the loop functions
$H_\mathrm{BP}$, $H_\mathrm{MP}$, $H_\mathrm{KR}$, $H_\mathrm{TD1}$, $H_\mathrm{TD2}$, and $
H_\mathrm{WF}$ are also given there.
It is interesting to note that although separately these loop functions are divergent (scale-dependent) and
some of them contain power-counting breaking pieces ($H_\mathrm{KR}$ and $H_\mathrm{MP}$), the overall
contributions are finite and do not break power-counting. This is an explicit manifestation of the Ademollo-Gatto theorem.

\begin{figure}[t]
\includegraphics[scale=0.35,angle=270]{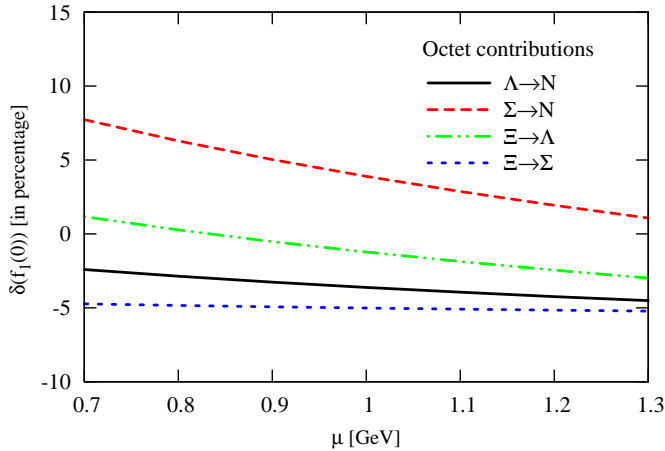}
\caption{Scale dependence of the octet contributions to the SU(3)-breaking corrections 
to the hyperon vector coupling $f_1(0)$.
\label{fig:omudep}}
\end{figure}

In the $\mathcal{O}(p^4)$ calculation, we have implemented mass-splitting
corrections in a similar way as  Ref.~\cite{Lacour:2007wm} except that we have used the masses  obtained from the second-order ChPT fit, as described above, instead of the physical masses.
Similar to the IRChPT study of Ref.~\cite{Lacour:2007wm}, the $\mathcal{O}(p^4)$ results
contain higher-order divergences. We have removed the infinities using the modified minimal-subtraction ($\overline{MS}$) scheme. The analytical results
are quite lengthy and will not be shown here. In Fig.~\ref{fig:omudep}, we show the scale dependence of the octet contributions, which is rather mild for most cases except for the $\Sigma\rightarrow N$ transition.  The scale dependence can be
used to estimate higher-order contributions by varying $\mu$ in a reasonable range. In the following, we present
the results by varying $\mu$ from
0.7 to 1.3 GeV.
It should be mentioned that if we had adopted the same method as Ref.~\cite{Villadoro:2006nj} to calculate the $\mathcal{O}(p^4)$ contributions, i.e.,
by expanding the results and keeping only those linear in baryon mass splittings, our 
$\mathcal{O}(p^4)$ results would have been convergent.  

We have checked that our results up to $\mathcal{O}(p^3)$ are the same as those obtained in Ref.~\cite{Krause:1990xc},
while in the $M_B\sim\Lambda_{\chi SB}$ limit our results recover the HBChPT ones~\cite{Villadoro:2006nj} at both $\mathcal{O}(p^3)$ and $\mathcal{O}(p^4)$ including
the $1/M$ recoil corrections. All these are known to explicitly verify the Ademollo-Gatto theorem in the sense of Eq.~(\ref{eq:ag}).

\begin{table}[htbp]
      \renewcommand{\arraystretch}{1.5}
     \setlength{\tabcolsep}{0.4cm}
     \caption{Values for the masses and couplings appearing in the calculation of
the SU(3)-breaking corrections to $f_1(0)$. \label{table:para}}
\begin{tabular}{cl|cl}
\hline\hline
 $D$& $0.8$ & $M_B$ & $1.151$ GeV \\
 $F$& $0.46$ & $M_D$ &$1.382$ GeV\\
 $f_\pi$ & $0.0924$ GeV & $M_0$ & $1.197$ GeV \\
 $F_0$& $1.17f_\pi$ & $b_D$ & $-0.0661$ GeV$^{-1}$\\
 $m_\pi$ & $0.138$  GeV&$b_F$ & $0.2087$ GeV$^{-1}$\\
 $m_K$ & $0.496$  GeV & $M_{D0}$ & $1.216$ GeV \\
 $m_\eta$ & $0.548$ GeV & $\gamma_M$ & $0.3236$ GeV$^{-1}$\\
 $\mathcal{C}$& 1.0\\
\hline\hline
\end{tabular}
\end{table}

Table \ref{table:octetfull} shows the SU(3)-breaking corrections in the notation of 
Eq.~(\ref{eq:Adem-Gatt}). For comparison, we
also list the numbers obtained in HBChPT~\cite{Villadoro:2006nj} and IRChPT~\cite{Lacour:2007wm}. The numerical values are obtained with
the parameters given in Table \ref{table:para}.  As in Ref.~\cite{Geng:2008mf} we have used an average $F_0=1.17 f_\pi$ with $f_\pi=92.4$ MeV. It should be pointed out that the HBChPT and the IRChPT
results are obtained using $f_\pi$. 

First, we note that in three of the
four cases, the $\delta^{(3)}$ numbers are smaller than the $\delta^{(2)}$ ones. The situation is
similar in IRChPT  but 
quite different in HBChPT. In
the HBChPT calculation~\cite{Villadoro:2006nj},
the $\delta^{(3)}$ contribution is larger than the $\delta^{(2)}$ one for the four cases.\footnote{What we denote by $\delta^{(3)}$  is the sum of
those labeled by $\alpha^{(3)}$ and $\alpha^{(1/M)}$ in Ref.~\cite{Villadoro:2006nj}.} 
This tells that recoil corrections (in the HBChPT language) or relativistic effects are important.
On the other hand, the results of the present work and those of IRChPT~\cite{Lacour:2007wm}, including the contributions of different chiral orders,
 are qualitatively similar. They are both very different from the HBChPT predictions, even for the signs in three of the
four cases. Obviously, as stressed in Ref.~\cite{Lacour:2007wm}, one
should trust more the relativistic  than the non-relativistic results, which have to be treated
with caution whenever $1/M$ recoil corrections become large.

\begin{table}[t]
      \renewcommand{\arraystretch}{1.5}
     \setlength{\tabcolsep}{0.4cm}
     \caption{Octet contributions to the SU(3)-breaking corrections to $f_1(0)$ (in percentage). The central values of the  $\mathcal{O}(p^4)$ results are calculated with $\mu=1$ GeV and the uncertainties are
obtained by varying $\mu$ from 0.7 to 1.3 GeV.\label{table:octetfull}}
\begin{tabular}{c|ccc|c|c}
\hline\hline
&\multicolumn{3}{c|} {present work}& HBChPT~\cite{Villadoro:2006nj} & IRChPT~\cite{Lacour:2007wm}\\\hline
                        & $\delta^{(2)}$&  $\delta^{(3)}$ & $\delta^{(2)}+\delta^{(3)}$& $\delta^{(2)}+\delta^{(3)}$& $\delta^{(2)}+\delta^{(3)}$\\
			\hline
 $\Lambda\rightarrow N$ & $-3.8$  &  $0.2^{+1.2}_{-0.9}$ & $-3.6^{+1.2}_{-0.9}$  & $2.7$ & $-5.7\pm2.1$ \\
 $\Sigma\rightarrow N$  & $-0.8$ &  $4.7^{+3.8}_{-2.8}$ & $3.9^{+3.8}_{-2.8}$   & $4.1$ & $2.8\pm0.2$\\
 $\Xi\rightarrow\Lambda$& $-2.9$  &  $1.7^{+2.4}_{-1.8}$ & $-1.2^{+2.4}_{-1.8}$ & $4.3$ & $-1.1\pm1.7$\\
 $\Xi\rightarrow\Sigma$ & $-3.7$  &  $-1.3^{+0.3}_{-0.2}$ & $-5.0^{+0.3}_{-0.2}$  & $0.9$ & $-5.6\pm1.6$\\
 \hline\hline
\end{tabular}
\end{table}
It is clear 
from Table \ref{table:octetfull} that the convergence is slow even in the case of the relativistic calculations, a well known feature of
SU(3) baryon ChPT. It is then necessary to have a way to calculate ``higher-order''
contributions. Going to $\mathcal{O}(p^5)$ one needs to introduce
unknown LEC's such that the predictive power of ChPT is lost. An alternative approach is 
to consider the contributions of dynamical heavier resonances. A basic assumption of ChPT is
 that these heavier degrees of freedom can be integrated out with
their effects incorporated in the LEC's.
However, that may not be totally true in the one-baryon sector since the gap between the lowest baryon octet
and the lowest baryon decuplet is only $\sim 0.3$ GeV, not very different from the pion mass and even
smaller than the kaon(eta) mass. Therefore, it is necessary to investigate their contributions.
In the HBChPT scheme, this task has recently been 
undertaken in Ref.~\cite{Villadoro:2006nj}, where it is concluded that the decuplet contributions
completely spoil the convergence of the chiral series.
We study the contributions of the decuplet baryons in the covariant framework
in the following section.

\subsection{SU(3)-breaking corrections to $f_1(0)$ induced by dynamical decuplet baryons up to $\mathcal{O}(p^4)$}
\begin{figure}[t]
\includegraphics[scale=0.8,angle=270]{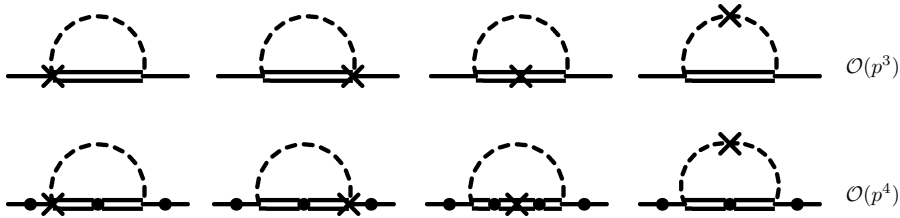}
\caption{Feynman diagrams contributing to the leading and next-to-leading order SU(3)-breaking corrections 
to the hyperon vector coupling $f_1(0)$, through dynamical decuplet baryons. The 
notations are the same as those of Fig.~\ref{fig:doctet} except that double lines
indicate decuplet baryons. We have not shown explicitly those diagrams corresponding to wave function renormalization, which have been included in the calculation.
\label{fig:ddecup}}
\end{figure}

Fig.~\ref{fig:ddecup} shows the diagrams that contribute 
to SU(3)-breaking corrections to  $f_1(0)$ with
dynamical decuplet baryons up to $\mathcal{O}(p^4)$.  It should be noted
that unlike in the HBChPT case~\cite{Villadoro:2006nj}, Kroll-Rudermann (KR) kind of diagrams also contribute.
In fact, using the consistent coupling scheme of Ref.~\cite{Pascalutsa:2000kd}, there are four
KR diagrams: Two are from minimal substitution in the derivative of the pseudoscalar fields and
 the other 
two are from minimal substitution in the derivative of the decuplet fields (see 
Eq.~(\ref{eq:consistent}) and also Ref.~\cite{Geng:2009hh}).

The $\mathcal{O}(p^3)$ results are relatively simple and have the following general structure for the transition
$i\rightarrow j$:
\begin{eqnarray}\label{eq:sumd3}
 \delta_D^{(2)}(i\rightarrow j)&=&\sum_{M=\pi,\eta,K}\gamma^\mathrm{BP}_M D_\mathrm{BP}(m_M)+\sum_{M=\pi,\eta} \gamma^\mathrm{MP}_{M} D_\mathrm{MP}(m_M,m_K)+\sum_{M=\pi,\eta,K} \gamma^\mathrm{KR}_M D_\mathrm{KR}(m_M)\nonumber\\
&&+\frac{1}{2}\sum_{M=\pi,\eta,K}(\gamma^\mathrm{WF}_M(i)+\gamma^\mathrm{WF}_M(j))D_\mathrm{WF}(m_M),
\end{eqnarray}
where $\gamma^\mathrm{BP}$, $\gamma^\mathrm{MP}$, $\gamma^\mathrm{KR}$, and $\gamma^\mathrm{WF}$ are
listed in Tables \ref{table:d3bp}, \ref{table:d3mp}, \ref{table:d3kr}, and \ref{table:d3wf} of the Appendix. The loop
functions $D^\mathrm{BP}$, $D^\mathrm{MP}$, $D^\mathrm{KR}$, and $D^\mathrm{WF}$ can be calculated
analytically, but they are quite lengthy. In the Appendix, they are given in terms of Feynman-parameter integrals,
which can be easily integrated.

\begin{figure}[b]
\includegraphics[scale=0.35,angle=270]{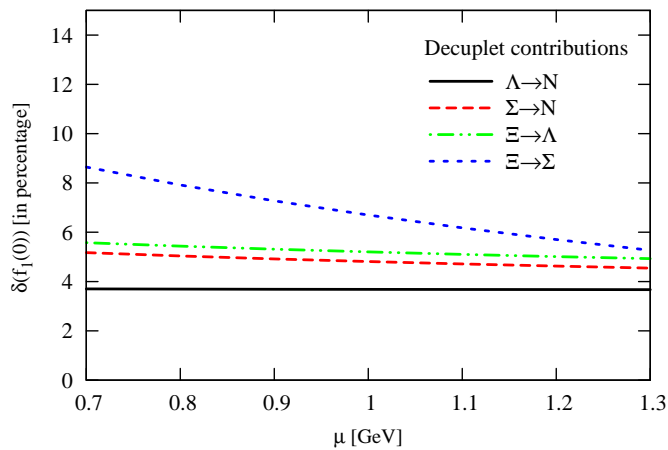}
\caption{Scale dependence of the decuplet contributions to the SU(3)-breaking corrections 
to the hyperon vector coupling $f_1(0)$.
\label{fig:dmudep}}
\end{figure}

To calculate the $\mathcal{O}(p^4)$ chiral contributions, we implement the decuplet-baryon mass splittings in the same
way as in the octet case. The $\mathcal{O}(p^4)$ results contain again higher-order divergences, with
the scale dependence shown in Fig.~\ref{fig:dmudep}. The infinities have been removed by
the $\overline{MS}$ procedure. The dependence is found to be rather mild except
for the $\Xi\rightarrow\Sigma$ transition. In this case, unlike in the octet case, the divergences
cannot be removed by expanding and keeping only terms linear in baryon and decuplet mass splittings.
The full $\mathcal{O}(p^4)$ analytical results are quite involved and, therefore, they will not be shown here.

The numerical results obtained with the parameter values given in Table \ref{table:para}
are summarized in Table \ref{table:decupfullpas}. It can be seen that at $\mathcal{O}(p^3)$, the decuplet contributions are relatively small compared to the octet ones at the same order.  On the other hand, the $\mathcal{O}(p^4)$ contributions are sizable and all of them have positive signs. 

Using the conventional Lagrangians of Ref.~\cite{Hacker:2005fh}, one 
obtains different numbers and different $\mu$ dependence. In the heavy-baryon limit, however, the results obtained with both coupling schemes are 
found to be the same and convergent, confirming the fact that the differences induced by the ``consistency'' procedure are of higher chiral order~\cite{Pascalutsa:2000kd} (see also Ref.~\cite{Geng:2009hh}).

\begin{table}[t]
      \renewcommand{\arraystretch}{1.5}
     \setlength{\tabcolsep}{0.4cm}
     \caption{Decuplet contributions to the SU(3)-breaking corrections to $f_1(0)$ (in percentage). The central values of the $\mathcal{O}(p^4)$ result are calculated with $\mu=1$ GeV and the uncertainties are obtained by varying $\mu$ from 0.7 GeV to 1.3 GeV.
     \label{table:decupfullpas}}
\begin{tabular}{c|ccc|ccc}
\hline\hline
&\multicolumn{3}{c}{Present work}&\multicolumn{3}{c}{HBChPT}\\\hline
& $\delta^{(2)}$&  $\delta^{(3)}$ & $\delta^{(2)}+\delta^{(3)}$  & $\delta^{(2)}$&  $\delta^{(3)}$ & $\delta^{(2)}+\delta^{(3)}$\\
			\hline
 $\Lambda\rightarrow N$ &    $0.7$ & $3.0^{+0.1}_{-0.1}$ & $3.7^{+0.1}_{-0.1}$ & $1.8$ & $1.3$ & $3.1$ \\
 $\Sigma\rightarrow N$  & $-1.4$ & $6.2^{+0.4}_{-0.3}$ & $4.8^{+0.4}_{-0.3}$& $-3.6$ & $8.8$ & $5.2$\\
 $\Xi\rightarrow\Lambda$&  $-0.02$ & $5.2^{+0.4}_{-0.3}$ & $5.2^{+0.4}_{-0.3}$ &$-0.05$ & $4.2$ & $4.1$ \\
 $\Xi\rightarrow\Sigma$ &  $0.7$ & $6.0^{+1.9}_{-1.4}$ & $6.7^{+1.9}_{-1.4}$ &$1.9$ & $-0.2$ & $1.7$\\
 \hline\hline
\end{tabular}
\end{table}
\begin{table}[t]
      \renewcommand{\arraystretch}{1.5}
     \setlength{\tabcolsep}{0.4cm}
     \caption{SU(3)-breaking corrections to $f_1(0)$ up to $\mathcal{O}(p^4)$ (in percentage), including both
the octet and the decuplet contributions.
     \label{table:full}}
\begin{tabular}{c|ccc}
\hline\hline
& $\delta^{(2)}$&  $\delta^{(3)}$ & $\delta^{(2)}+\delta^{(3)}$ \\
			\hline
 $\Lambda\rightarrow N$ &    $-3.1$ & $3.2^{+1.3}_{-1.0}$ & $0.1^{+1.3}_{-1.0}$ \\
 $\Sigma\rightarrow N$  & $-2.2$ & $10.9^{+4.2}_{-3.1}$ & $8.7^{+4.2}_{-3.1}$\\
 $\Xi\rightarrow\Lambda$&  $-2.9$ & $6.9^{+2.8}_{-2.1}$ & $4.0^{+2.8}_{-2.1}$\\
 $\Xi\rightarrow\Sigma$ &  $-3.0$ & $4.7^{+2.2}_{-1.6}$ & $1.7^{+2.2}_{-1.6}$ \\
 \hline\hline
\end{tabular}
\end{table}
 
In Table \ref{table:decupfullpas}, the numbers denoted by HBChPT differ from those of Ref.~\cite{Villadoro:2006nj}.
The $\delta^{(2)}$ column would have coincided if we had used the same values for the couplings $\mathcal{C}=0.8$ and  $F_0=0.0933$ GeV. On the other hand, our $\delta^{(3)}$ contributions due to the octet baryon mass splittings are much smaller than those of Ref.~~\cite{Villadoro:2006nj}. It is interesting to note that unlike in the octet case, the
HBChPT results are similar to the relativistic ones.\footnote{The HB results are obtained in a slightly different way than
the relativistic ones. To obtain the $\mathcal{O}(p^3)$ numbers, we have used $M_B=1.151$ GeV and $M_D=1.382$ GeV and have performed an expansion in terms of
the decuplet-octet mass splitting, $M_D-M_B$. To obtain the $\mathcal{O}(p^4)$ ones, we have used physical masses
for both the octet and the decuplet baryons and have performed an additional expansion keeping only
the terms linear in the
octet and the decuplet baryon mass splittings. Although this procedure is the same as that  of Ref.~\cite{Villadoro:2006nj}, we get different results. We find that the discrepancy comes from the octet mass-splitting corrections to the meson-pole diagram of Fig.~3. If we had mistakenly exchanged
the masses of the mesons in the loop, we would have obtained the same results as those of Ref.~\cite{Villadoro:2006nj}.}

\begin{figure}[htpb]
\includegraphics[scale=0.35,angle=270]{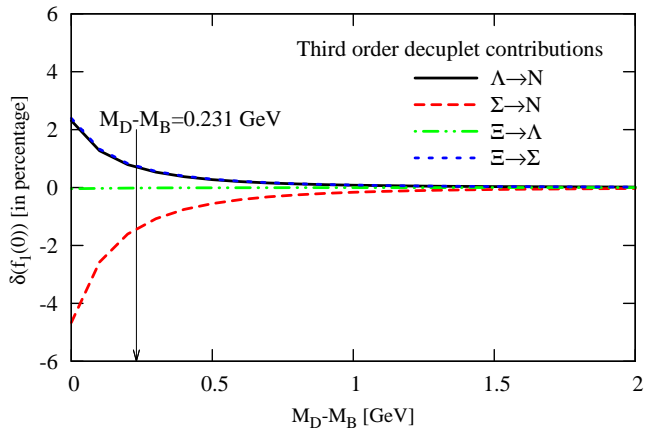}
\caption{Decuplet $\mathcal{O}(p^3)$ contributions to the SU(3)-breaking corrections 
to the hyperon vector coupling $f_1(0)$ as a function of the decuplet-octet mass splitting
$M_D-M_B$.
\label{fig:decoup}}
\end{figure}

As the decuplet-octet mass splitting increases,
one expects that the decuplet contributions decrease and eventually vanish as
the splitting goes to infinity. This is indeed the case, as can be clearly seen from  Fig.~\ref{fig:decoup}, where 
the $\mathcal{O}(p^3)$ decuplet contributions are plotted as a function of the decuplet-octet mass splitting.

\subsection{Full results and comparison with other approaches}

Summing the octet  and the decuplet contributions, we obtain
the numbers shown in Table \ref{table:full}. Two things are noteworthy. First, the convergence 
is slow, even taking into account the scale dependence of the $\delta^{(3)}$ corrections. Second, for three of the four transitions, the
$\delta^{(3)}$ corrections have a different sign than the $\delta^{(2)}$ ones.

In Table \ref{table:comparison}, we compare our results with those obtained from other approaches,
including large $N_c$ fits~\cite{FloresMendieta:2004sk}, quark models~\cite{Donoghue:1986th,Schlumpf:1994fb, Faessler:2008ix}, and two quenched LQCD calculations~\cite{Guadagnoli:2006gj,Sasaki:2008ha}. The large $N_c$ results in general favor positive corrections, which are consistent with our central values. Two of
the quark models predict negative corrections, while that of Ref.~\cite{Faessler:2008ix}
favors positive corrections. It is interesting to note that in Ref.~\cite{Faessler:2008ix}
the valence quark effects give negative contributions, as in the other two quark models, while the chiral
effects provide positive contributions, resulting in net positive corrections. Our numbers also agree, within uncertainties, with the quenched LQCD ones. In principle, LQCD calculations provide another model-independent way
to obtain the SU(3)-breaking corrections to $f_1(0)$. At present, however, the quenched LQCD calculations are not yet accurate enough to determine these numbers, due to the large quark masses used in the simulation and other systematic uncertainties. 
\begin{table}[b]
      \renewcommand{\arraystretch}{1.5}
     \setlength{\tabcolsep}{0.4cm}
     \caption{SU(3)-breaking corrections (in percentage) to $f_1(0)$ obtained
in different approaches.
     \label{table:comparison}}
\begin{tabular}{c|c|c|ccc|c}
\hline\hline
& Present work & Large $N_c$ & \multicolumn{3}{c|}{Quark model} & quenched LQCD \\\hline
& &Ref.~~\cite{FloresMendieta:2004sk}   & Ref.~\cite{Donoghue:1986th} &Ref.~\cite{Schlumpf:1994fb} & Ref.~\cite{Faessler:2008ix} &  \\\hline
 $\Lambda\rightarrow N$ &    $0.1^{+1.3}_{-1.0}$ & $2\pm2$ & $-1.3$ & $-2.4$ & $0.1$ & \\
 $\Sigma\rightarrow N$  & $8.7^{+4.2}_{-3.1}$ &  $4\pm3$ &  $-1.3$ & $-2.4$ & $0.9$ & $-1.2\pm2.9\pm4.0$~\cite{Guadagnoli:2006gj}\\
 $\Xi\rightarrow\Lambda$ &  $4.0^{+2.8}_{-2.1}$&  $4\pm4$ &  $-1.3$ & $-2.4$ & $2.2$ & \\
 $\Xi\rightarrow\Sigma$ &  $1.7^{+2.2}_{-1.6}$ &  $8\pm5$ &  $-1.3$ & $-2.4$ & $4.2$ & $-1.3\pm1.9$~\cite{Sasaki:2008ha}\\
 \hline\hline
\end{tabular}
\end{table}

Finally, we will briefly discuss the implications of our results 
for the estimation of $V_{us}$. There have been several previous attempts
to extract this parameter using hyperon semileptonic decays  
\cite{Cabibbo:2003cu,Cabibbo:2003ea,FloresMendieta:2004sk,Mateu:2005wi}.
As discussed in Ref.~\cite{Mateu:2005wi} a rather clean determination of 
$f_1 V_{us}$ can be done by using $g_1/f_1$ and the decay rates from experiment
and taking for $g_2$ and $f_2$ their SU(3) values. This latter approximation
is supported by the fact that their contributions to the decay rate are
reduced by kinematic factors (See, for instance, Eq. (10) of Ref.~\cite{FloresMendieta:2004sk}).
Using the values of $f_1 V_{us}$ compiled in Table 3 of Ref.~\cite{Mateu:2005wi}
and our results for $f_1$ we get
\begin{equation}
\label{eq:vus}
V_{us} =0.2177\pm 0.0030\,.
\end{equation}
This value is consistent with the large $N_c$ fits of Ref.~\cite{FloresMendieta:2004sk} 
and with the result obtained from $\tau$ decays\cite{Gamiz:2007qs}, and lower
than the results of kaon decays~\cite{Blucher:2005dc} 
and the fits to hyperon decays from Refs.~\cite{Cabibbo:2003cu,Cabibbo:2003ea}.
Although the quoted error seems competitive with calculations using other processes,
we must remark that the error estimation for Eq. (\ref{eq:vus}) includes only the experimental errors and the 
uncertainties related to the scale dependence. Other systematic uncertainty sources, like the effect
of higher order SU(3)-breaking corrections are hard to estimate and have not been included.
Even with these limitations, our results clearly point to  positive values for the SU(3)-breaking corrections to
$f_1$ and therefore towards relatively small values of $V_{us}$.

\section{Summary and conclusions}
We have performed a study of the SU(3)-breaking corrections to the hyperon vector coupling $f_1(0)$ in covariant baryon chiral perturbation theory including both the octet and the decuplet contributions. We confirm earlier findings in HBChPT and
IRChPT that the convergence of the chiral series is slow in the case with only
dynamical octet baryons. Our study of the decuplet contributions shows that at $\mathcal{O}(p^3)$  they are in general smaller than those of their octet counterparts, while at $\mathcal{O}(p^4)$ they are sizable.
Combining both octet and decuplet contributions, we found positive SU(3)-breaking corrections
to all the four independent $f_1(0)$'s, which compare favorably with the large $N_c$ fits and those of
the quark model taking into account chiral effects.

The fact that the $\mathcal{O}(p^4)$ chiral contributions are comparable to the
$\mathcal{O}(p^3)$ ones suggests that the $\mathcal{O}(p^5)$ chiral effects
may not be negligible. We have estimated their size by varying $\mu$ from
0.7 to 1.3 GeV.  Taking into account these higher-order uncertainties, our results still
favor positive SU(3)-breaking corrections to the four $f_1(0)$'s.

An  accurate determination of $V_{us}$ from hyperon semileptonic decays depends largely
on our knowledge on the value of $f_1(0)$. While the SU(3)-symmetric values have been used
in some fits to extract $V_{us}$,  most studies have taken into account
SU(3)-breaking corrections to  $f_1(0)$. We have provided the first covariant baryon
ChPT predictions for $f_1(0)$ up to $\mathcal{O}(p^4)$ including both the octet and the decuplet contributions. We encourage their
uses in new attempts to extract $V_{us}$ from  hyperon decay data. 

\section{Acknowledgments}
This work was partially supported by the  MEC grant  FIS2006-03438 and the European Community-Research Infrastructure
Integrating Activity Study of Strongly Interacting Matter (Hadron-Physics2, Grant Agreement 227431) under the Seventh Framework Programme of EU. L.S.G. acknowledges support from the MICINN in the Program 
``Juan de la Cierva.'' J.M.C. acknowledges the same institution for a FPU grant.

\clearpage
\section{Appendix}
\subsection{Octet $\mathcal{O}(p^3)$ contributions}
In this subsection, we present the coefficients and loop functions
appearing in the calculation of the $\mathcal{O}(p^3)$ octet contributions, i.e., Eq.~(\ref{eq:sumo3}).
\begin{equation}
H_\mathrm{BP}=\frac{1}{(4\pi F_0)^2}\left[-3 m^2 -\frac{4 \cos^{-1}(\frac{m}{2M_B}) m^3}{\sqrt{4 M_B^2-m^2}}\left(\frac{m^2}{M_B^2}-3\right)+\frac{2 \log \left(\frac{m^2}{M_B^2}\right) m^4}{M_B^2}+2 \log \left(\frac{M_B^2 \mu }{m^3}\right) m^2\right],
\end{equation}

\begin{eqnarray}
 H_\mathrm{MP}&=&\frac{1}{(4\pi F_0)^2}\frac{1}{4 \left(m_1^2-m_2^2\right)}\times\nonumber\\
&&\Bigg[\frac{8 \left(\frac{m_1^2}{M_B^2}-4\right)\cos^{-1}\left(\frac{m_1}{2 M_B}\right) m_1^6 }{\sqrt{4 m_1^2 M_B^2-m_1^4}}-\frac{4 \log
   \left(\frac{m_1^2}{M_B^2}\right) m_1^6}{M_B^2}
+\bigg(6 \log \left(\frac{m_1^2}{M_B^2}\right) +2 \log \left(\frac{\mu^2}{M_B^2}\right) +11\bigg) m_1^4\nonumber\\
&&-\frac{8  \left(\frac{m_2^2}{M_B^2}-4\right) \cos^{-1}\left(\frac{m_2}{2M_B}\right)m_2^6}{\sqrt{4 m_2^2 M_B^2-m_2^4}}+\frac{4  \log \left(\frac{m_2^2}{M_B^2}\right)m_2^6}{M_B^2}
-(6
   \log \left(\frac{m_2^2}{M_B^2}\right)+2 \log \left(\frac{\mu^2}{M_B^2}\right)+11\bigg) m_2^4\nonumber\\
&&
+8 M_B^2 (1+  \log \left(\frac{\mu^2 }{M_B^2}\right))(m_1^2-
   m_2^2)\Bigg],
\end{eqnarray}

\begin{equation}
 H_\mathrm{KR}=\frac{1}{(4\pi F_0)^2}\left[\frac{\log \left(\frac{M_B}{m}\right) m^4}{M_B^2}-\frac{\sqrt{4 M_B^2-m^2} \cos^{-1}\left(\frac{m}{2 M_B}\right) m^3}{M_B^2}+
m^2( \log
   \left(\frac{\mu^2 }{M_B^2}\right) +2)+M_B^2(1+ \log \left(\frac{\mu^2 }{M_B^2}\right))\right],
\end{equation}

\begin{equation}
H_\mathrm{TD1}=\frac{1}{(4\pi F_0)^2}\left[\frac{ \log \left(\frac{\mu^2 }{m_1^2}\right) m_1^4-\log
   \left(\frac{\mu^2 }{m_2^2}\right)m_2^4}{m_1^2- m_2^2}+\frac{3}{2}(m_1^2+m_2^2)\right],
\end{equation}

\begin{equation}
 H_\mathrm{TD2}=\frac{1}{(4\pi F_0)^2}\left[(\log \left(\frac{\mu ^2}{m^2}\right) +1)m^2\right],
\end{equation}

\begin{equation}
 H_\mathrm{WF}=\frac{1}{(4\pi F_0)^2}\left[\frac{2 \log \left(\frac{m^2}{M_B^2}\right) m^4}{M_B^2}+\frac{4 \left(3-\frac{m^2}{M_B^2}\right) \cos^{-1}\left(\frac{m}{2 M_B}\right) m^3}{\sqrt{4
   M_B^2-m^2}}+\bigg(\log \left(\frac{\mu ^2}{m^2}\right)-2 \log \left(\frac{m^2}{M_B^2}\right) -3\bigg) m^2\right].
\end{equation}
\begin{table}[htbp]
      \renewcommand{\arraystretch}{1.5}
     \setlength{\tabcolsep}{0.4cm}
     \caption{Coefficients $\beta^\mathrm{BP}$ appearing in Eq.~(\ref{eq:sumo3}).\label{table:o3bp}}
\begin{tabular}{c|ccc}
\hline\hline
 Channel & $\pi$ loop & $\eta$ loop & $K$ loop \\ \hline
 $\Lambda\rightarrow N$& $-\frac{1}{2} D (D+F)$ & $-\frac{1}{6} D (D-3 F)$ & $-\frac{1}{6} (D-3 F)^2$ \\
 $\Sigma\rightarrow N$& $-\frac{1}{2} \left(D^2+3 F D+2 F^2\right)$ & $\frac{1}{6} D (D-3 F)$ & $-\frac{1}{2} (D+F)^2$ \\
 $\Xi\rightarrow \Lambda$& $\frac{1}{2} D (F-D)$ & $-\frac{1}{6} D (D+3 F)$ & $-\frac{1}{6} (D+3 F)^2$ \\
$\Xi\rightarrow\Sigma$ & $-\frac{1}{2} \left(D^2-3 F D+2 F^2\right)$ & $\frac{1}{6} D (D+3 F)$ & $-\frac{1}{2} (D-F)^2$\\
\hline\hline
\end{tabular}
\end{table}

\begin{table}[htpb]
      \renewcommand{\arraystretch}{1.5}
     \setlength{\tabcolsep}{0.4cm}
     \caption{Coefficients $\beta^\mathrm{MP}$ appearing in Eq.~(\ref{eq:sumo3}).\label{table:o3mp}}
\begin{tabular}{c|cc}
\hline\hline
 Channel & $\pi K$ loop & $\eta K$ loop   \\ \hline
 $\Lambda\rightarrow N$ & $\frac{1}{4} \left(3 D^2+2 F D+3 F^2\right)$ & $\frac{1}{12} (D+3 F)^2$ \\
 $\Sigma\rightarrow N$  & $\frac{1}{12} \left(D^2-18 F D+9 F^2\right)$ & $\frac{3}{4} (D-F)^2$ \\
 $\Xi\rightarrow\Lambda$& $\frac{1}{4} \left(3 D^2-2 F D+3 F^2\right)$ & $\frac{1}{12} (D-3 F)^2$ \\
 $\Xi\rightarrow \Sigma$& $\frac{1}{12} \left(D^2+18 F D+9 F^2\right)$ & $\frac{3}{4} (D+F)^2$\\
\hline\hline
\end{tabular}
\end{table}

\begin{table}[htbp]
      \renewcommand{\arraystretch}{1.5}
     \setlength{\tabcolsep}{0.4cm}
     \caption{Coefficients $\beta^\mathrm{KR}$ appearing in Eq.~(\ref{eq:sumo3}).\label{table:o3kr}}
\begin{tabular}{c|ccc}
\hline\hline
 Channel & $\pi$ loop & $\eta$ loop & $K$ loop \\\hline
 $\Lambda\rightarrow N$ &$-\frac{1}{4} \left(3 D^2+2 F D+3 F^2\right)$ & $-\frac{1}{12} (D+3 F)^2$ & $-\frac{1}{6} \left(5 D^2+6 F D+9 F^2\right)$ \\
 $\Sigma\rightarrow N$ & $-\frac{1}{12} \left(D^2-18 F D+9 F^2\right)$ & $-\frac{3}{4} (D-F)^2$ & $-\frac{1}{6} (D-3 F) (5 D-3 F)$ \\
 $\Xi\rightarrow\Lambda$ & $-\frac{1}{4} \left(3 D^2-2 F D+3 F^2\right)$ & $-\frac{1}{12} (D-3 F)^2$ & $-\frac{1}{6}(5D^2-6F D+9 F^2)$ \\
 $\Xi\rightarrow\Sigma$ & $-\frac{1}{12} \left(D^2+18 F D+9 F^2\right)$ & $-\frac{3}{4} (D+F)^2$ & $-\frac{1}{6} (D+3 F) (5 D+3 F)$\\
\hline\hline
\end{tabular}
\end{table}

\begin{table}[htbp]
      \renewcommand{\arraystretch}{1.5}
     \setlength{\tabcolsep}{0.4cm}
     \caption{Coefficients $\beta^\mathrm{WF}$ appearing in Eq.~(\ref{eq:sumo3}).\label{table:o3wf}}
\begin{tabular}{c|ccc}
\hline\hline
   & $\pi$ loop & $\eta$ loop & $K$ loop \\\hline
 $\Lambda$ & $ D^2$ & $\frac{D^2}{3}$ & $\frac{1}{3} \left(D^2+9 F^2\right)$ \\
 $\Sigma$ & $\frac{1}{3} \left(D^2+6 F^2\right)$ & $\frac{D^2}{3}$ & $D^2+F^2$ \\
 $N$        & $\frac{3}{4} (D+F)^2$ & $\frac{1}{12} (D-3 F)^2$ & $\frac{1}{6} \left(5 D^2-6 F D+9 F^2\right)$ \\
 $\Xi$ & $\frac{3}{4} (D-F)^2$ & $\frac{1}{12} (D+3 F)^2$ & $\frac{1}{6}(5 D^2+6F D+9 F^2)$ \\
\hline\hline
\end{tabular}
\end{table}

\clearpage
\subsection{Decuplet $\mathcal{O}(p^3)$ contributions}
In this subsection, we provide the coefficients and loop functions appearing
in the calculation of the $\mathcal{O}(p^3)$ decuplet contributions, i.e., Eq.~(\ref{eq:sumd3}).

\begin{eqnarray}
D_\mathrm{BP}&=&-\frac{\mathcal{C}^2}{(4\pi F_0)^2 M_D^2}\int\limits^1_0 dx\,M_B^2 (1-x) \left(\left((2 x-1) M_D^2+2 M_B x M_D-\left((x-2) M_B^2+2 m^2\right) x\right)\times\right.\nonumber\\
&&\hspace{3cm}\left. \log \left(\frac{\mu ^2}{\left((x-1)
   M_B^2+m^2\right) x-M_D^2 (x-1)}\right)-\left(M_B^2+2 M_D M_B+M_D^2-m^2\right) x\right),
\end{eqnarray}

\begin{eqnarray}
 D_\mathrm{MP}&=&-\frac{\mathcal{C}^2}{(4\pi F_0)^2 M_D^2}\int\limits^1_0 dx\,\int\limits^{1-x}_0 dy\,M_B^2 \left(2 M_B x (x M_B-M_B-M_D)+\left(-3 (x-1) x M_B^2+2 M_D x M_B-M_D^2 x-m_1^2 y\right.\right.\nonumber\\
  &&\left.\left.\hspace{4cm}    +m_2^2
   (x+y-1)\right) \log \left(\frac{\mu ^2}{(x-1) x M_B^2+M_D^2 x+m_1^2 y-m_2^2 (x+y-1)}\right)\right),
\end{eqnarray}

\begin{eqnarray}
 D_{KR}&=&-\frac{\mathcal{C}^2}{(4\pi F_0)^2 M_D^2}\int\limits^1_0 dx\,M_B (M_D+M_B x) \left(\left((x-1) M_B^2+m^2\right) x-M_D^2 (x-1)\right)\times\nonumber\\
 &&\hspace{3cm} \log \left(-\frac{\mu ^2}{M_D^2
   (x-1)-\left((x-1) M_B^2+m^2\right) x}\right),
\end{eqnarray}

\begin{eqnarray}
 D_\mathrm{WF}&=&-\frac{\mathcal{C}^2}{(4\pi F_0)^2 M_D^2}\int\limits^1_0 dx\, M_B \left(2 M_B^2 (x-1) x (M_B x-M_B-M_D)+\left(-5 M_B^3 (x-1)^2 x+4 M_B^2 M_D (x-1) x+\right.\right.\\
&&\left.\left.3 M_B (x-1)
   \left(m^2 (x-1)-M_D^2 x\right)+2 M_D \left(M_D^2 x-m^2 (x-1)\right)\right) \log \left(-\frac{\mu ^2}{m^2 (x-1)-x
   \left(M_B^2 (x-1)+M_D^2\right)}\right)\right).\nonumber
\end{eqnarray}

\begin{table}[htpb]
      \renewcommand{\arraystretch}{1.5}
     \setlength{\tabcolsep}{0.4cm}
     \caption{Coefficients $\gamma^\mathrm{BP}$ appearing in Eq.~(\ref{eq:sumd3}).\label{table:d3bp}}
\begin{tabular}{c|ccc}
\hline\hline
 Channel  & $\pi$ loop & $\eta$ loop & $K$ loop \\\hline
 $\Lambda\rightarrow N$   &  $-4 $ & $0$ & $-2 $ \\
 $\Sigma\rightarrow N$    & $-\frac{4 }{3}$ & $0$ & $-\frac{2 }{3}$ \\
 $\Xi\rightarrow \Lambda$ & $-2 $ & $0$ & $-2 $ \\
 $\Xi\rightarrow \Sigma$  & $-\frac{4 }{3}$ & $-2 $ & $-\frac{14 }{3}$\\
\hline\hline
\end{tabular}
\end{table}

\begin{table}[t]
      \renewcommand{\arraystretch}{1.5}
     \setlength{\tabcolsep}{0.4cm}
     \caption{Coefficients $\gamma^\mathrm{MP}$ appearing in Eq.~(\ref{eq:sumd3}).\label{table:d3mp}}
\begin{tabular}{c|cc}
\hline\hline
 Channel & $\pi K$ loop  & $\eta K$ loop   \\ \hline
 $\Lambda\rightarrow N$  & $1$ & $0$ \\
 $\Sigma\rightarrow N$   & $-2 $ & $-1$ \\
 $\Xi\rightarrow\Lambda$ & $0$ & $-1$ \\
 $\Xi\rightarrow\Sigma$  & $1$ & $2 $\\
\hline\hline
\end{tabular}
\end{table}

\begin{table}[t]
      \renewcommand{\arraystretch}{1.5}
     \setlength{\tabcolsep}{0.4cm}
     \caption{Coefficients $\gamma^\mathrm{KR}$ appearing in Eq.~(\ref{eq:sumd3}).\label{table:d3kr}}
\begin{tabular}{c|ccc}
\hline\hline
  Channel                      & $\pi$ loop & $\eta$ loop & $K$ loop \\\hline
 $\Lambda\rightarrow N$& $7 $ & $0$ & $3 $ \\
 $\Sigma\rightarrow N$ & $\frac{14}{3}$ & $1$ & $\frac{13}{3}$ \\
 $\Xi\rightarrow\Sigma$& $4 $ & $1$ & $5 $ \\
 $\Xi\rightarrow\Sigma$& $\frac{5}{3}$ & $2$ & $\frac{19}{3}$\\
\hline\hline
\end{tabular}
\end{table}

\begin{table}[t]
      \renewcommand{\arraystretch}{1.5}
     \setlength{\tabcolsep}{0.4cm}
     \caption{Coefficients $\gamma^\mathrm{WF}$ appearing in Eq.~(\ref{eq:sumd3}).\label{table:d3wf}}
\begin{tabular}{c|ccc}
\hline\hline
     & $\pi$ loop & $\eta$ loop & $K$ loop \\\hline
 $\Lambda$   & $-3 $ & $0$ & $-2 $ \\
 $\Sigma$   & $-\frac{2 }{3}$ & $-1$ & $-\frac{10}{3}$ \\
 $N$          & $-4 $ & $0$ & $-1$ \\
 $\Xi$      & $-1$ & $-1$ & $-3$ \\
\hline\hline
\end{tabular}
\end{table}
\end{document}